\documentclass[twocolumn,aps,prc,superscriptaddress,showpacs]{revtex4}
\usepackage{amsmath,bm}
\usepackage{graphicx}
\usepackage{dcolumn}
\usepackage{multirow}
\newcolumntype{z}[1]{D{.}{.}{#1}}
\begin{document}
\draft

\title{Phenomenological Scaling of Rapidity Dependence for Anisotropic Flows
in 25 MeV/nucleon Ca + Ca by Quantum Molecular Dynamics Model }

\author{T. Z. Yan }
\affiliation{Shanghai Institute of Applied Physics, Chinese
Academy of Sciences, Shanghai 201800, China} \affiliation{Graduate
School of the Chinese Academy of Sciences, Beijing 100080, China}

\author{Y. G. Ma }
 \thanks{ Corresponding author. Email: ygma@sinap.ac.cn }

\author{X. Z. Cai}
\author{D. Q. Fang}
\author{G. C. Lu}
\author{W. Q. Shen}
\author{W. D. Tian}
\author{H. W. Wang}
\author{K. Wang}

\affiliation{Shanghai Institute of Applied Physics, Chinese
Academy of Sciences, Shanghai 201800, China}

\date{\today}

\begin{abstract}

Anisotropic flows ($v_1$, $v_2$, $v_3$ and $v_4$) of light
fragments up till the mass number 4 as a function of rapidity have
been studied for 25 MeV/nucleon $^{40}$Ca + $^{40}$Ca at large
impact parameters by Quantum Molecular Dynamics model. A
phenomenological scaling behavior of rapidity dependent flow
parameters $v_n$ (n = 1, 2, 3 and 4) has been found as a function
of mass number plus a constant term, which may arise from the
interplay of collective and random motions. In addition,
$v_4/{v_2}^2$ keeps almost independent of rapidity and remains a
rough constant of 1/2 for all light fragments.

\end{abstract}


\maketitle

Collective flow is an important observable in heavy ion collisions
and it can bring some information on nuclear equation of state and
in-medium nucleon-nucleon cross section
\cite{Olli,Voloshin,Sorge,Danile,Zhang,Li,Shu,Kolb,Zheng,Gale,yg-prc,INDRA,Ko,Chen,Chen-PRL}.
In intermediate energy heavy ion collisions, flows can reflect the
interplay of collective and random motions. For a thermalized
system, the random motions of emitted fragments are dictated by
thermal energy, which is independent of mass. In contrast,
fragment energy increases linearly with mass  if there exists
collective motion \cite{Huang,Partlan,Jeong}. Many studies of the
properties of the directed flow ($v_1$) and the elliptic flow
($v_2$) have been carried out and much interesting physics has
been demonstrated on the properties and origin of the collective
motion in both nucleonic or partonic levels. Recently, we carried
out a Quantum Molecular Dynamics model calculation and found that
there is a nucleon-number scaling for the transverse momentum
dependent  elliptic flow for light particles up to the mass number
A = 4 \cite{Yan_PLB,Ma_AIP,Ma_NPA}. But there are little studies
about higher order flows such as $v_3$ and $v_4$ as well as their
rapidity dependence. In this work, we shall present rapidity
dependence of $v_1$, $v_2$, $v_3$ and $v_4$ flows for a symmetric
collision, namely $^{40}$Ca + $^{40}$Ca collisions at 25
MeV/nucleon and large impact parameters ($b$ = 4-6 fm) simulated
by Quantum Molecular Dynamics  model. The scaling behaviors of
$v_4/(v_2)^2$  as a function of rapidity is also presented.

Anisotropic flow is defined as the different $n$-th harmonic
coefficient $v_n$ of the Fourier expansion for the particle
invariant azimuthal distribution \cite{Voloshin}
\begin{equation}
\frac{dN}{d\phi}\propto{1+2\sum^\infty_{n=1}{v_n\cos(n\phi)}},
\end{equation}
where $\phi$ is the azimuthal angle between the transverse
momentum of the particle and the reaction plane. Note that the
z-axis is defined as the direction along the beam and the impact
parameter axis which is labelled as x-axis. Anisotropic flows
generally depend on both particle transverse momentum and
rapidity, and for a given rapidity the anisotropic flows at
transverse momentum $p_t$ ($p_t = \sqrt{p_x^2+p_y^2}$) can be
evaluated according to
\begin{equation}
v_{n}(p_t) = \langle cos(n\phi) \rangle,
\end{equation}
where $\langle \cdots \rangle$ denotes average over the azimuthal
distribution of particles with transverse momentum $p_t$. The
anisotropic flows $v_n$ can further be expressed in term of
single-particle averages:
\begin{equation}
v_1 = \langle cos\phi \rangle = \langle \frac{p_x}{p_t} \rangle,
\end{equation}
\begin{equation}
 v_2 = \langle cos(2\phi) \rangle = \langle
\frac{p^2_x-p^2_y}{p^2_t} \rangle,
\end{equation}
\begin{equation}
 v_3 = \langle cos(3\phi) \rangle = \langle
\frac{p^3_x-3p_{x}p^2_y}{p^3_t} \rangle,
\end{equation}
\begin{equation}
 v_4 = \langle cos(4\phi) \rangle = \langle
\frac{p^4_x-6p^2_{x}p^2_y+p^4_y}{p^4_t} \rangle,
\end{equation}
where $p_x$ and $p_y$ are, respectively, projections of particle
transverse momentum in and perpendicular to the reaction plane.

The intermediate energy heavy-ion collision dynamics is complex
since both mean field and nucleon-nucleon collisions play the
competition roles. Furthermore, the isospin dependent role should
be also incorporated for asymmetric reaction systems. Isospin
dependent Quantum Molecular Dynamics model (IDQMD) has been
affiliated with isospin degrees of freedom with mean field and
nucleon-nucleon collision \cite{J. Aichelin,Y. G. Ma1,H. Y.
Zhang,Y. B. Wei,Y. G. Ma3}. The IDQMD model can explicitly
represent the many body state of the system and principally
contain correlation effects to all orders and all fluctuations,
and can describe the time evolution of the colliding system well.
When the spatial distance ($\Delta r$) is closer than 3.5 fm and
the momentum difference ($\Delta p$) is smaller than 300 MeV/c
between two nucleons, two nucleons can coalesce into a cluster
\cite{J. Aichelin}. With this simple coalescence mechanism which
has been extensively applied in transport theory, different size
clusters can be recognized.

In the model the nuclear mean-field potential is parameterized as
\begin{equation}
U(\rho,\tau_{z}) = \alpha(\frac{\rho}{\rho_{0}}) +
\beta(\frac{\rho}{\rho_{0}})^{\gamma} +
\frac{1}{2}(1-\tau_{z})V_{c} \nonumber
\end{equation}
\begin{equation}
+ C_{sym} \frac{(\rho_{n} - \rho_{p})}{\rho_{0}}\tau_{z} + U^{Yuk}
\end{equation}
where $\rho_0$ is the normal nuclear matter density
($0.16fm^{-3}$), $\rho_n$, $\rho_p$ and $\rho$  are the  neutron,
proton and total densities, respectively. $\tau_z$ is $z$-th
component of the isospin degree of freedom, which equals 1 or -1
for neutrons or protons, respectively. The coefficients $\alpha$,
$\beta$ and $\gamma$ are parameters for nuclear equation of state.
$C_{sym}$ is the symmetry energy strength due to the density
difference of neutrons and protons in nuclear medium, which is
important for asymmetry nuclear matter
(here $C_{sym} = 32$ MeV is used), but it is trivial for the
symmetric system studied in the present work. $V_c$ is the Coulomb
potential and $U^{Yuk}$ is Yukawa (surface) potential. In this
work, we take $\alpha $ = 124 MeV, $\beta$ = 70.5 MeV and $\gamma$
= 2 which corresponds to the so-called hard EOS with an
incompressibility of $K$ = 380 MeV.

\begin{figure}
\vspace{-0.1truein}
\includegraphics[scale=0.3]{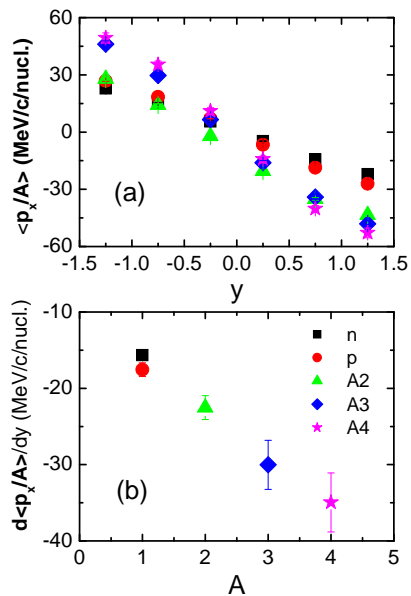}
\vspace{-0.1truein}
 \caption{\footnotesize (a) The average in-plane transverse momentum per nucleon as a function of
the normalized rapidity in the c.m. frame. (b) Mass dependence of
$d \langle p_x/A \rangle/dy$. Squares represent for neutrons,
circles for protons, triangles for fragments of A=2, diamonds for
A=3 and stars for A=4. Error bars show the errors of the fits.
}\label{dpxdy}
\end{figure}

 About 60000 events have been simulated for the collision system
of $^{40}$Ca + $^{40}$Ca at 25 MeV/nucleon with impact parameter
from 4 fm to 6 fm. The physics results were extracted at the time
of 200 fm/c when the system has been in the freeze-out stage.
Fig.~\ref{dpxdy}(a) shows rapidity dependence of the average
in-plane transverse momentum per nucleon $\langle p_x/A \rangle$
around mid-rapidity for light fragments. Here rapidity $y$ is the
rapidity of fragment in the center of momentum (c.m.) frame which
is normalized to the initial projectile rapidity, i.e., $y =
y_{c.m.}/y_{proj}$. It shows that $\langle p_x/A \rangle$
decreases monotonously with the increasing rapidity, and decreases
more rapidly for heavier fragments. The negative slope is mainly
driven by the attractive mean field in this energy region
\cite{Wil,yg-prc}. Fig.~\ref{dpxdy}(b) shows mass dependence of
the so-called sidewards flow parameter (slope value) $d \langle
p_x/A \rangle /dy$, which can be easily extracted via linear fits
near mid-rapidity in Fig.~\ref{dpxdy}(a). The flow parameter is
negative and its absolute value increases with mass apparently,
which seems consistent with coalescence-like cluster production
mechanism \cite{Huang}.

\begin{figure}
\vspace{-0.1truein}
 \includegraphics[scale=0.8]{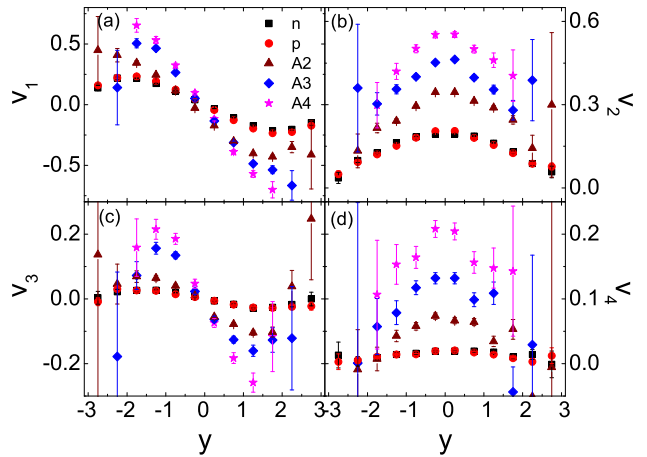}
\vspace{-0.1truein}
 \caption{\footnotesize (a), (b), (c), (d) show rapidity dependence of
 $v_1$, $v_2$, $v_3$, $v_4$ for light fragments respectively.
 Squares represent for neutrons, circles for protons, triangles for
fragments of A=2, diamonds for A=3 and stars for A =
4.}\label{v-y}
\end{figure}

In Fig.~\ref{v-y}, we show integrated anisotropic flows $v_1$,
$v_2$, $v_3$ and $v_4$ as a function of rapidity $y$. The trend of
the directed flow $v_1$ shown in Fig.~\ref{v-y}(a) is similar to
that of $\langle p_x/A \rangle $, i.e., the directed flow
decreases monotonously with the increasing rapidity around
mid-rapidity. It is positive in target-like rapidity region and
negative in projectile-like rapidity region. And it gets to the
extremum near $y=\pm1.5$, and then the absolute value of directed
flow goes down with the increasing $\mid y \mid$. It also shows at
a given rapidity the absolute magnitude of $v_1$ is larger for
heavier clusters. Fig.~\ref{v-y}(b) presents the rapidity
dependence of elliptic flow $v_2$. It shows that the flow is
positive and the curve for a given fragment is symmetric with
$y=0$, and descends monotonously when the $\mid y \mid$ increases.
Again, heavier fragments have larger elliptic flow at the same
rapidity. The rapidity dependences of $v_1$ and $v_2$ are similar
to the results at RHIC energies  for charged hadrons
\cite{J.Adams}, but the interaction level of the matter in the two
different energy region are obviously different. In this low
energy heavy ion collisions, the positive elliptic $v_2$
essentially stems from the collective rotational motion
\cite{Wil,yg-prc}. Fig.~\ref{v-y}(c) and (d) show rapidity
dependence of higher order flows of $v_3$ and $v_4$, respectively.
 The trend of $v_3$ and $v_4$ are similar to that of $v_1$
and $v_2$, respectively, but they show smaller magnitude of flows
for a given particle in comparison with $v_1$ and $v_2$,
respectively. Nevertheless the magnitude of flows of $v_3$ and
$v_4$ are not too small and hence higher order flows should not be
neglected in this energy region.

\begin{figure}
\vspace{-1.5truein}
\includegraphics[scale=0.42]{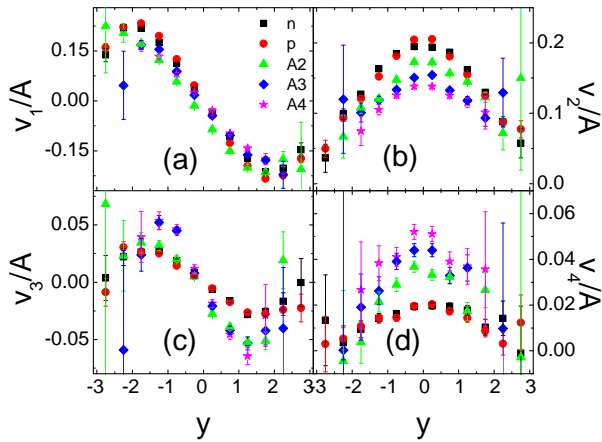}
\vspace{-1.truein}
 \caption{\footnotesize (a), (b), (c) and (d) show rapidity dependence of
 $v_1$, $v_2$, $v_3$ and $v_4$ per nucleon for light fragments, respectively.
 Squares represent for neutrons, circles for protons, triangles for
fragments of A=2, diamonds for A=3 and stars for A=4.}\label{va-y}
\end{figure}

\begin{figure}
\vspace{-1.3truein}
\includegraphics[scale=0.42]{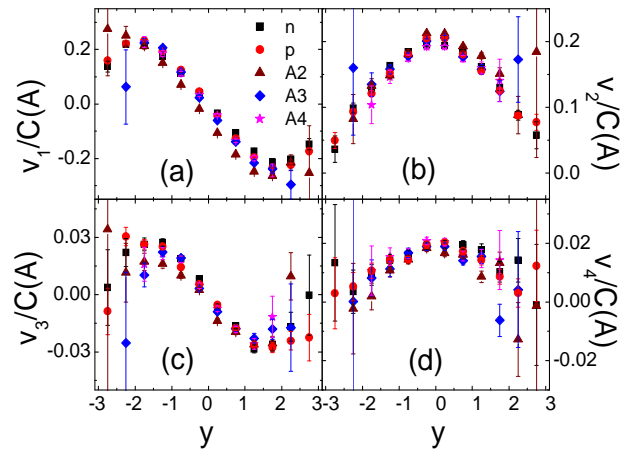}
\vspace{-1.2truein}
 \caption{\footnotesize Same as Fig.~\ref{va-y}, but for $v_n/C(A)$ (n=1, 2, 3, 4) versus rapidity.}\label{vc-y}
\end{figure}

For testing if the  number of nucleon scaling works for rapidity
dependent flows, we plot the flows per nucleon as a function of
rapidity which is shown in Fig.~\ref{va-y}. It seems that the
curves for different fragments do not stay together as we imagined
except for $v_1/A$. One  reason is that the effect of random
motion which is independent of mass may weaken the mass scaling in
rapidity space. For details, if we compare the magnitude of
collective flows of other fragments with that of proton or
neutron, it shows the flow magnitude of fragments with mass number
of 2, 3, and 4 is smaller than proton or neutron for $v_1$ and
$v_2$, but larger than proton or neutron for $v_3$ and $v_4$.
However, if we use a mass number dependent function $C(A) =
\frac{5}{8}(A+\frac{3}{5})$ instead of $A$ for $v_1$ and $v_2$,
and use $C(A) = 3(A-\frac{2}{3})$ for $v_3$ and $v_4$, and plot
$v_n/C(A)$ versus $y$ again which is shown in Fig.~\ref{vc-y}, we
can now see the curves for different fragments coincide with each
others excellently except for slight deviation in large rapidity
region, i.e. in the spectators region. The coefficient $C(A)$ may
be qualitatively understood by two parts: the constant part
reflects the effect of random motion which is independent of mass,
and the part including mass number $A$ reflects the collective
motion which increases linearly with mass \cite{Huang,Kunde}. From
the mass dependent coefficient of $C(A)$, there are two classes:
one for $v_1$ and $v_2$, and another for $v_3$ and $v_4$. The
larger coefficient for $v_3$ and $v_4$ may indicate that the
contribution of collective motion to higher order flows $v_3$ and
$v_4$ is larger than that to $v_1$ and $v_2$.

\begin{figure}
 \vspace{-1.1truein}
\includegraphics[scale=0.35]{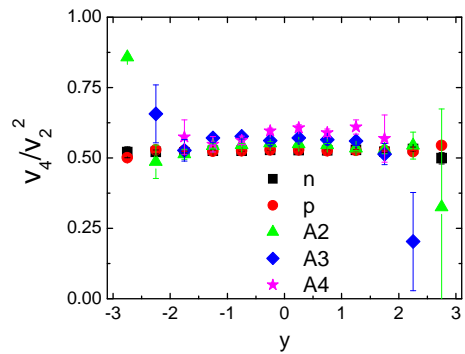}
\vspace{-1.truein}
 \caption{\footnotesize The ratio of $v_4/v^2_2$ for
neutrons (squares), proton (circles), the fragments of $A=2$
(triangles), $A=3$ (diamonds) and  $A=4$ (stars) versus rapidity
y.}\label{v4v22}
\end{figure}

The RHIC experimental data demonstrated a scaling relationship
between 2nd flow $v_2$ and $n$-th flow $v_n$ for hadrons, namely
$v_n(p_t)\sim v^{n/2}_2(p_t)$ and $v_4/v_2^2 \sim 1.2$ in the STAR
data \cite{STAR03}. And we have already shown the scaling of
$v_4/v^2_2$  as a function of $p_t$ in intermediate energy heavy
ion collisions \cite{Yan_PLB,Ma_AIP,Ma_NPA}, i.e., the ratio of
$v_4/v^2_2$ is independent of transverse momentum and shows a
constant value of $0.5$ for different light fragments. Similarly,
we plot $p_t$-integrated $v_4/v^2_2$ as a function of rapidity
which is shown in Fig. ~\ref{v4v22}. The figure shows that the
ratios of $v_4/v^2_2$ for different fragments up to $A = 4$
roughly keep  a constant of $0.5$ in the studied rapidity region
except for some fluctuations at large rapidities. Therefore the
ratio of $v_4/v^2_2$ is almost independent of transverse momentum
and rapidity and its value is around 0.5. In the frame of
relativistic fluid dynamics which works in RHIC energies, the
ratio value of 0.5 reflects that the collision system reaches to
thermal equilibrium and its subsequent evolution follows the laws
of ideal fluid dynamics \cite{Bro}. A coincident value of 0.5 as
relativistic fluid dynamics in the present intermediate energy may
also indicate some kinds of thermal equilibrium has been also
reached.

To summarize, phenomenological behaviors of anisotropic flows as a
function of rapidity for light fragments up till mass number 4
have been investigated by the simulation of 25 MeV/nucleon
$^{40}$Ca + $^{40}$Ca collision in peripheral collisions in the
framework of the Quantum Molecular Dynamics model. It was shown
that $v_1$ and $v_3$ of light fragments decrease monotonously with
rapidity from positive value near target-like rapidity to negative
value near projectile-like rapidity, while $v_2$ and $v_4$ are
positive and show Gaussian-like shape with a peak around
$y_{c.m.}$ = 0. When we plot anisotropic flows per nucleon
($v_1/A$, $v_2/A$, $v_3/A$ and $v_4/A$) versus rapidity for light
fragments, the curves of different particles do not stay together
perfectly except for $v_1/A$. But when we plot $v_n/C(A)$ (n = 1,
2, 3, 4) versus rapidity where $C(A)$ is a linear function of the
mass number $A$ plus an additional constant term, the curves of
different particles collapse onto the same curve. That indicates
that the fragment flows may arise from the interplay of collective
(a term proportional to mass number in $C(A)$) and random thermal
motion (a constant term in $C(A)$) of nucleons. $C(A)$ shows a
classification between $v_{1,2}$ and $v_{3,4}$, this might reflect
that the different role of collective motion on the different
anisotropies in momentum space. Additionally, it was found that
the ratio of $v_4/v^2_2$ are almost independent of rapidity and
the value is about $0.5$, which implies a thermalization scenario
of the ideal fluid-like dynamics could be applied even in such low
energy heavy ion collisions. Further theoretical investigations
and experimental checks are awaiting.

The work was supported partially by the National Basic Research
Program of China (973 Program) under Contract No. 2007CB815004,
Shanghai Development Foundation from Science and Technology under
Grant Numbers 06JC14082 and 06QA14062, the National Natural
Science Foundation of China under Grant No 10535010 and 10775167.
{}
\end{document}